\documentclass
[twocolumn,showpacs,aps,amsmath,amssymb,superscriptaddress]{revtex4}%
\usepackage{amssymb,amscd}
\usepackage[T1]{fontenc}
\usepackage{graphicx}
\usepackage{amsmath}
\usepackage{amsfonts}
\usepackage{amssymb}%
\usepackage{natbib}

\providecommand{\U}[1]{\protect\rule{.1in}{.1in}}
\begin{document}
\title{Conductance in Co|Al$_{2}$O$_{3}$|Si|Al$_{2}$O$_{3}
$ Permalloy with asymmetrically doped barrier}
\author{R. Guerrero}
\author{F. G. Aliev}
\author{R. Villar}
\affiliation{Dpto. de Fisica de la Materia Condensada, C-III, Universidad Autonoma de
Madrid, 28049, Madrid, Spain}
\author{T. Santos}
\author{J. Moodera}
\affiliation{Francis Bitter Magnet Laboratory, Massachusetts Institute of Technology,
Cambridge, MA 02139, USA}
\author{V.~K.~Dugaev}
\affiliation{Department of Physics, Rzesz\'ow University of Technology, al. Powsta\'nc\'ow
Warszawy 6, 35-959 Rzesz\'ow, Poland, and Department of Physics and CFIF,
Instituto Superior T\'ecnico, TU Lisbon, av. Rovisco Pais, 1049-001 Lisbon, Portugal}
\author{J. Barna\'s}
\affiliation{Department of Physics, Adam Mickiewicz University,
Umultowska~85, 61-614 Pozna\'n, Poland; and Institute of Molecular
Physics, Polish Academy of Sciences, Smoluchowskiego 17, 60-197
Pozna\'n, Poland}
\date{\today}

\begin{abstract}
We report on dependence of conductance and tunnelling magnetoresistance on
bias voltage at different temperatures down to 2K in Co%
$\vert$%
Al$_{2}$O$_{3}$(10\AA )%
$\vert$%
Si($\delta$)%
$\vert$%
Al$_{2}$O$_{3}$(2\AA )%
$\vert$%
Permalloy magnetic tunnel junctions. Complementary low frequency noise
measurements are used to understand the conductance results. The obtained data
indicate the breakdown of the Coulomb blockade for thickness of the asymmetric
silicon layer exceeding 1.2\AA . The crossover in the conductance, the
dependence of the tunnelling magnetoresistance with the bias voltage and the
noise below 80K correspond to 1 monolayer coverage. Interestingly, the zero
bias magnetoresistance remains nearly unaffected by the presence of the
silicon layer. The proposed model uses Larkin-Matveev approximation of
tunnelling through a single impurity layer generalized to 3D case and takes into
account the variation of the barrier shape with the bias voltage. The main
difference is the localization of all the impurity levels within a single
atomic layer. In the high thickness case, up to 1.8\AA , we have introduced a
phenomenological parameter, which reflects the number of single levels on the
total density of silicon atoms.

\end{abstract}
\maketitle


\section{Introduction}

The discovery of large tunnelling magnetoresistance (TMR) at room temperature
\cite{Moodera95,Miyazaki95} has strongly renewed the interest in spin
tunnelling phenomena. Up to very recently the main efforts were concentrated
on the increase of tunnelling magnetoresistance values by using ferromagnetic
electrodes with the highest possible spin polarization (half metallic
ferromagnetism), searching for new types of insulating barriers (including the
so called spin filters \cite{leclairEuS02}), or a combination of both
approaches, where the ferromagnetic/insulator interface design could also play
an important role for spin polarized tunnelling. The last approach has
recently provided an enormous progress in the TMR values at room temperatures.
It has been demonstrated that in epitaxial Fe/MgO/Fe magnetic tunnel junctions
(MTJ), where there exist conditions for coherent propagation of specific spin
orbitals from one ferromagnetic electrode to the other one, TMR has reached
experimentally values up to 410$\%$ \cite{yuasa2006gtm}. This fact, supported
by theoretical calculations predicting a TMR of more than 1000$\%$ have
further increased both the fundamental and technological interest in spin
polarized tunnelling.

Another possible research direction, which remains however poorly explored, is
related to tunnelling in complex (hybrid) junctions. Indeed, the manipulation
of the barrier by doping with magnetic or nonmagnetic impurities, or inserting
magnetic, for example see reference \cite{CBFert1997}, nonmagnetic
\cite{CB2003} (even superconducting \cite{takahashi1999sia,CB2003}) quantum
dots (QD), would add a new degree of freedom to spin polarized tunnelling and
strongly enhance the versatility of spintronic devices based on spin polarized
tunnelling. Tunnelling in such hybrid, but non magnetic devices, has been
intensively studied during the last two decades and especially in single
electron transistors where the gate electrode is attached to quantum dots
separated by two barriers each with a metallic contact (emitter and
collector)\cite{maekawa2006sdt}.

Recent theoretical studies of hybrid spintronic devices with two ferromagnetic
leads contacting single or double quantum dots have revealed plenty of new
interesting phenomena related with the interplay between magnetic tunnelling
processes and spin/charge accumulation on quantum dots. From the experimental
side, some few groups have demonstrated that Coulomb interaction may indeed
play an important role not only in ferromagnetic granular systems
\cite{CB2002} (as occurs in the corresponding nonmagnetic analogs), but also
in ferromagnetic single electron tunnelling devices constructed either from a
2D electron gas \cite{CB1996} or when a single metallic nanoparticle is
contacted by ferromagnetic electrodes \cite{CBFert2006}. It has also been
reported that in ultrasmall nanoparticles, with a diameter between 5-10nm, their electronic structures (i.e.
quantum effects) may also influence spin polarized tunnelling
\cite{Deshmukh2002}. Other interesting examples of spin polarized tunnelling
in hybrid structures include ferromagnetic leads contacting carbon nanotubes
\cite{nanotubes2005} or a C60 molecule \cite{C60Martinek2004}.

Actually, from the technological point of view, it is easier to attach
ferromagnetic leads via tunnelling barriers to an array instead of to a single
quantum dot. This method, despite the evident drawback due to some
distribution in QD sizes and the corresponding charging energies, adds evident
versatility to the design of the experiment, allowing continuous tuning
between two different regimes: (i) weak doping regime where QDs are
substituted by impurities and (ii) strong doping regime. In the second regime
one would expect the QD charging energy to reduce continuously with the
average dot size, allowing sequential electron transport through an otherwise
blocked channel (Coulomb blockade) in addition to direct tunnelling.

Therefore, the knowledge of different mechanisms affecting the
conductance, and especially its dependence on the applied bias,
could be an important instrument for a comparative analysis of
noise and transport. We should, however, note that the bias
dependence of conductivity and TMR in real magnetic tunnel
junctions, is still poorly understood. The most accepted model
\cite{Slonczewski}, which does not take into account the possible
defects inside the barrier, predicts an attenuation of the
polarization due to the decrease of the difference between the
height of the barrier and the bias voltage. However, the observed
reduction in conductivity usually exceeds the predicted effect.

Two theories have been developed in order to explain the anomalous reduction
in TMR. The first one explores the existence of impurity states, which reduce
the spin current polarization and influence the conductance of the junctions
at low bias \cite{bratkovskybiasdep97}. The second one involves inelastic
tunnelling as the main origin of unpolarized current \cite{Zhangbiasdep97}.
This point of view has been also supported by the measurement of inelastic
tunnelling in magnetic tunnel junctions \cite{MooderaIETS98}. Later on Ding et
al \cite{dingZBAvacuum03} have also detected a tunnelling magnetoresistance in
vacuum based magnetic tunnel junctions with a reduced dependence on the bias
voltage. These results indicate a possible influence of the impurity states on
the polarization of the tunnel current.

For the nonmagnetic tunnel junctions with (nonmagnetic) nanoparticles inside
the barrier the presence of a zero bias anomaly was first reported and
explained by Giaever \cite{Giaever69}. The main mechanism responsible for the
appearance of the threshold voltage is Coulomb blockade, which controls two
steps tunnelling. In this model the threshold voltage is distributed from 0 V
to a maximum voltage $V_{s}$. In the doped tunnel junctions $V_{s}$ is given
by the size of the doping particles, which determines a charging energy, given
by the capacitance of the particles. This provides some distribution in the
population of electrons inside the particles.

The two steps tunnelling in a magnetic tunnel junction has been later treated
using the other method developed by Glazman and Matveev
\cite{Glazman88,Larkin87}. This theory states that the current is defined, in
each particle, by the tunnelling rates from one of the electrodes to the
central particle, and from the island to the other electrode. As soon
as conductance and spin current polarization are modified, a modification of
the dependence on voltage of the tunnelling magnetoresistance may be expected.
Indeed, when the bias voltage is increased, the number of allowed two steps
processes is also increased.

Previously to this work, Jansen et al. \cite{Jansen2000} studied MTJs with Si
nanoparticles up to 1.8\AA \ introduced in a symmetric position. They observed
gradual suppression of tunnelling magnetoresistance. In the present work Si
particles were introduced asymmetrically inside the barrier. While symmetric
doping effectively separates the barrier into two parts with similar tunneling
rates, the asymmetric doping is expected to affect weakly the largest
tunneling rate minimizing the influence of the nonmagnetic Si doping (at least
for relatively weak doping levels).

This work presents an experimental study of electron transport in Co(100\AA )/Al$_{2}$O$_{3}$(10\AA )/Si($\delta$)/Al$_{2}%
$O$_{3}$(2\AA )/Py(100\AA ) hybrid magnetic tunnel junctions, where the
largest barrier is 5 times larger than the short one. We observed a continuous
transition between the weak (Si impurities) and strong (array of Si quantum
dots) regimes. In order to discriminate the different conductance regimes, we
study the temperature and bias dependence of both the conductivity and TMR as
a function of Si doping.

\section{Experimental details.}

Details of sample preparation have been published previously \cite{Jansen2000}. For silicon doped samples, the tunnel barriers were deposited in two steps. After deposition of the underlying Co electrode, a first tunnel barrier was
formed by deposition and subsequent oxidation of 10~\AA \ of Al. Subsequently, sub-monolayer amounts of Si ($\delta$ in the following) were deposited on the Al$_{2}$O$_{3}$ surface, followed by a second Al layer deposition (2~\AA ) and oxidation, resulting in a ``$\delta$-doped'' Al$_{2}$O$_{3}$|Si|Al$_{2}$ O$_{3}$ tunnel barrier. After the deposition of the barrier 100~\AA \ of permalloy were deposited, in order to form the second magnetic layer. All the samples thicknesses were measured using a quartz monitor. The quality of the Al layers and the Al$_{2}$O$_{3}$ has been previously tested in Ref. \cite{mooderaAPL1997,Jansen2000II}. In the following the positive bias voltage corresponds to the application of the voltage in the top electrode, while the negative to application of the voltage in the bottom one.

Measurements were performed using a computer controlled system \cite{guerrero2005APL}, which allows
to detect the dynamic resistance, the DC value of the current and the voltage,
and the noise in the device under study. Biasing of the samples was done at a
constant current, applied using a calibrated source. It also allowed to
modulate the applied current. A square waveform was used in order to detect
the transfer function of the line and the dynamic resistance of the junctions.
The voltage response of the devices was amplified by using DC coupled low
noise amplifiers. The amplified signal was recorded in an analog-digital converter.

The measurement of the noise uses the same biasing technique and the same low
noise amplifiers, which were placed in the top part of a cryostat. The
pre-amplified signals are further amplified by additional low-noise amplifiers
(Stanford Research SR560). A spectrum analyzer SR780 calculates the
cross-correlation spectrum of the voltage noise, containing thermal, shot and
$1/f$ contributions. The obtained dynamic resistance allows to convert the
voltage noise into the current noise. Extrinsic noise, introduced by the
amplifiers and the current source, was removed by using the data extracted
from a careful calibration performed on resistors at low temperatures. In the experiments we measured 9 different  samples, 2 of each Si thickness, except for $\delta$=0.6\AA where we only charaterized 1.

\section{Conductance, zero bias anomaly and tunnelling magnetoresistance}

\begin{figure}[ptb]
\includegraphics[width=\linewidth,clip=]{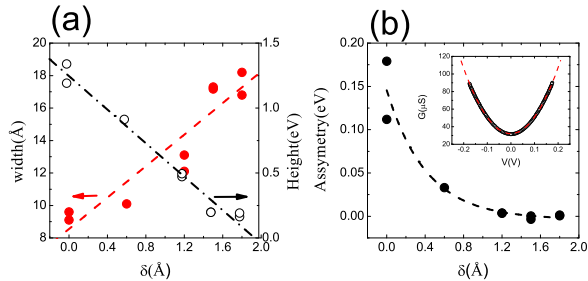}\caption{Barrier parameters
evaluated using the BDR model. The graph (a) shows the barrier width and height in the left and the right axis, respectively, as function of the Si thickness. In the graph (b) is shown the asymmetry also as function of the Si thickness. In the inset we present the typical dependence on voltage of the conductance at room temperature. The dashed line is a parabolic fit used to extract the parameters plotted in the graph (a) and (b). }%
\label{h&w}%
\end{figure}

At room temperature the dependence of the conductance on the bias voltage fits
well to a parabolic function (see inset to Fig. \ref{h&w}b). The parabolic
dependence of the conductance, being due to direct tunnelling, was explained
within the Brinkman-Dynes-Rowell (BDR) model \cite{brinkman}, which describes
the tunnelling conductance as a function of the parameters of a trapezoidal
barrier: the width, the average height and the difference between the sides of
the barrier, such difference being known as asymmetry.

The results of the fits are plotted in Fig. \ref{h&w}, showing in general an
increase of the barrier width with Si doping. However, the increase of the
width does not correspond with the deposited Si thickness. As to the height of
the barrier, it clearly diminishes when the silicon is introduced inside the
barrier. At present we have no clear explanation for this reduction. One of
the possible reasons could be some decrease of the work function of the
aluminium oxide. This explanation, however, contradicts the observed variation
of the barrier asymmetry (Fig. \ref{h&w}), which decreases with the thickness
of the Si layer. The disagreement between the deposited thickness and the obtained parameters using the BDR model could be attributed either to the simplification made by the model, which assums a trapezoidal barrier, a parabolic band structure and the WKB approximation,	or  to defects inside the barrier that should diminish the barrier height. However the obtained results present evidence that the Si layer affects the barrier properties beyond the errors committed in the estimation procedure.

\begin{figure}[t]
\includegraphics[width=0.8\linewidth,clip=]{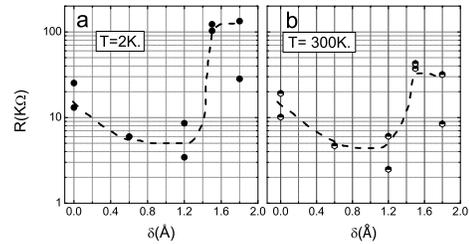}\caption{Resistance of the
MTJ's at low temperature, graph (a) and  room temperature, graph (a),  as a
function of the Si thickness ($\delta$(\AA )). The lines in the graphs are guides to the eyes.}%
\label{res300K&2K}%
\end{figure}

The tunnelling resistance of the studied Si doped MTJs measured at $T=$2 K and
$T=$300 K, in the zero bias limit, is plotted in Fig. \ref{res300K&2K}. The
observed enhancement of the resistance when the temperature is lowered rules
out the presence of pinholes even for the highest silicon thickness \cite{rudiger2001,jonsson-akerman2000}. There is
an increase of the resistance for $\delta\geq$1.5~\AA , both at room and low
temperatures, whereas at lower concentrations the resistance trend is to
decrease. This fact indicates a change of the regime in the conductance, which
will be further confirmed in the dependence of the conductance on the bias
voltage and the TMR.

\subsection{Dependence of the conductance on the voltage at low temperatures}

While at room temperature conductance is a parabolic function of the bias, at
low temperatures (below 100~K), the dependence of conductance on the bias
changes substantially. On the one hand, all MTJs at low biases (V$\lesssim
$30~mV) present a peak in the resistance. Usually such a peak is called zero
bias anomaly (ZBA). On the other hand, high bias conductance regimes (above
100~mV) show a strong variation with silicon thickness.

To show more clearly the qualitative change of conductance regime with Si
doping we present the bias dependence of the normalized conductance
\begin{equation}
\text{ZBA}(\%)=100\times\frac{R(100\text{mV}-R(0\text{mV})} {R(100\text{mV})}
\end{equation}

\begin{figure}[b]
\includegraphics[width=0.8\linewidth,clip=]{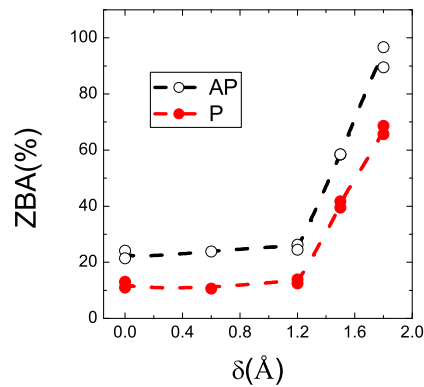}\caption{Dependence of the
quantity $ZBA$ (defined in the text) on silicon thickness at $T$=2 K in the parallel (P) and the antiparallel (AP) magnetic state. The
increased $ZBA$ at higher doping layers reflects a different behavior, related
with the formation of Si islands. }
\label{ZBA}%
\end{figure}

The dependence of ZBA vs. Si thickness, plotted in Fig. \ref{ZBA}, shows that
the resistance peak, being weakly dependent on $\delta$ for low Si thickness,
strongly increases for the high doping region. The crossover region
corresponds to the thickness of approximately $\delta$=1.2~\AA .

\begin{figure}[b]
\includegraphics[width=\linewidth,clip=]{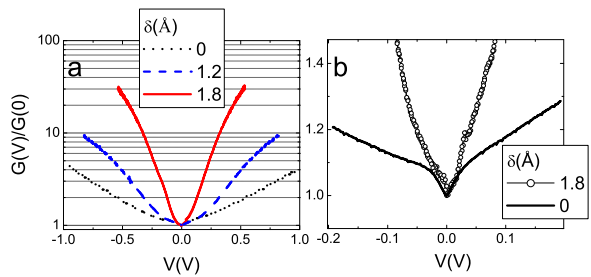}\caption{Dependence of the
conductance on the bias voltage at $T$=2 K. The curves are normalized by the
conductance measured at 0mV in each curve. The plot (a) presents data up to
high bias, where a mechanism related to the formation of Si islands
becomes important (see the text). The graph (b) presents the low bias
behavior, showing the presence of the same zero bias anomaly at V$\lesssim
$30mV.}
\label{GV}%
\end{figure}

The experimental data presented above may be understood within the two-step
model as follows. If the formation of silicon islands starts for a Si
thickness $\delta$=1.2~\AA , then the enhanced resistance peak (ZBA) could be
attributed to the appearance of a new energy scale in the electron transport
through the barrier, related with the finite electron capacitance of the
silicon islands, being practically absent for the small Si thickness range.
This hypothesis is supported by the plot of the normalized bias voltage
dependence of the conductance, Fig. \ref{GV}. Clearly in the conductance of the junctions with higher Si doping levels there is a crossover at a certain voltage. At higher voltage the dependence of the conductance becomes less pronounced, whereas the junctions with low doping maintain the same behavior, as expected in a non-doped tunnel junction with electron transport due to direct tunneling.

The change of the conductance regime is also evident from comparison of the
lower bias conductance for undoped MTJs and those with highest Si doping
(corresponding to $\delta$=1.8 \AA ). As can be seen in the (b) graph in Fig.
\ref{GV}, both curves show structure in G(V) at low bias close to V=30~mV. The
similarity of these weak anomalies both for the undoped and doped MTJs
indicate their common origin, most probably related with electron conduction
mechanisms through the aluminium oxide barrier.

\begin{figure}[ptb]
\includegraphics[width=0.8\linewidth,clip=]{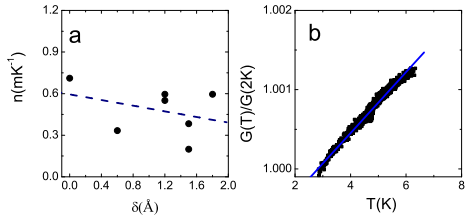}\caption{The graph (a) plots the slope of the temperature dependence of the conductance
normalized by its value at the lowest temperature.The graph (b) shows the
low temperature linear dependence of the conductance on temperature.}
\label{LTslope}
\end{figure}

We have observed that generally, for all MTJs studied, the low bias
conductance varies linearly with temperature at low temperatures (T<20~K).
Within the Coulomb blockade model this could be attributed to a variation in
the thermally activated population of electrons inside the islands
\cite{wolf}, which determines the slope of the conductance at zero bias. In
brief, the conductance at zero bias and low temperatures could be expressed as

\begin{equation}
G(0,T)\propto\int_{0}^{\infty}n\left(  V_{Ch}\right)  e^{-\frac{eV_{Ch}}%
{k_{B}T}}dV_{Ch}\sim n\left(  V_{Ch}\sim0\right)  \frac{k_{B}}{e}T
\end{equation}

where $eV_{Ch}$ is the charging energy needed to introduce an
electron in the metallic layer. At $V=$0~V the number of charged
islands is given by the exponential term. Then, at low
temperatures, the conductance is proportional to the number of
charged particles in thermal equilibrium $n\left(  0\right) $.
Although the temperature dependence of conductivity of the
different MTJs studied varies in nearly two orders of magnitude,
the normalized (to 2~K) low temperature slope in the linear
dependence of conductivity vs. temperature was found to be much
weakly dependent on the silicon thickness, with an average value
of $\left(  5\pm3\right)  \times10^{-4}$~K$^{-1}$ but with a
rather large dispersion (Fig. \ref{LTslope}).

\subsection{Tunneling magnetoresistance}

\begin{figure}[t]
\includegraphics[width=0.8\linewidth,clip=]{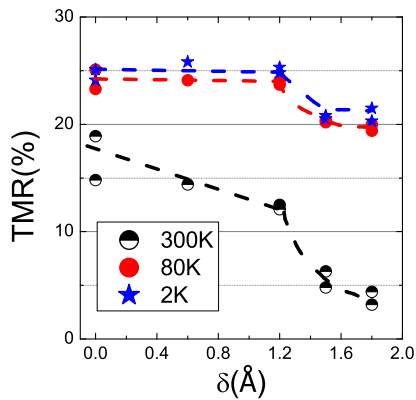}\caption{Dependence of the
tunnelling magnetoresistance on the silicon thickness ($\delta$(\AA )) at
three different temperatures. While presence of silicon reduces significantly the
polarization at room temperature, decreasing the value of the TMR ratio, for
temperatures below 80K the presence of the silicon layer seems not to affect
the spin polarization, for $\delta\leq$1.2\AA \.{ . }Some small influence
appears at $\delta>1.2$~\AA . }%
\label{TMRTemp}%
\end{figure}

The dependence of the tunnelling magnetoresistance on Si thickness is shown in
figure \ref{TMRTemp}. This plot represents the zero bias tunnelling
magnetoresistance, obtained by using the following definition of TMR%

\begin{equation}
\text{TMR(\%)}=100\times\frac{R_{AP}-R_{P}} {R_{P}}
\end{equation}
where $R_{AP}$ and $R_{P}$ are the resistance in the antiparallel
and the parallel states, respectively.

We have analyzed zero bias TMR vs. Si thickness for three different
temperatures (300 K, 80 K, 2 K). The influence of the silicon doping on TMR is
strongest at room temperature, suppressing TMR in nearly one order of
magnitude for the highest Si thickness (1.8~\AA ). The low temperature TMR
values, however, remain nearly unaffected by the silicon. The step-like
reduction of TMR at low temperatures (of about 10$\%$ at 77 K and 2 K) was
observed for a Si thickness of $\delta\geq$1.2~\AA . This apparent reduction
of the tunnelling magnetoresistance may be directly linked to variation of the
ZBA with Si content (shown in Fig. \ref{ZBA}).

Indeed, for small Si thickness, only direct tunnelling is possible, due to
Coulomb blockade. This weakly changes TMR at low temperatures for small bias,
which does not activate possible spin mixing due to the spin flip processes
introduced by the Si. The Coulomb blockade is suppressed for the doping range
of $\delta\geq$1.5~\AA , as indicated by conductance vs. voltage measurements
(Fig. \ref{GV}), opening new conductance channels related to two-step
tunnelling via the array of Si dots. The newly opened conductance channels
create also a source of unpolarized current due to spin mixing and loose of
spin memory of the electrons tunnelling through the array of Si dots.
Suppression of Coulomb blockade just for the Si thickness bigger than 1.2{\AA} could be due to activation of a segregation process of Si
atoms to nanometer scale dots, acting as a real capacitance. This is contrary
to the behaviour in low Si thickness junctions,  where seems reasonable to suppose that Si could be more homogenously diluted inside the Al$_{2}$O$_{3}$ barrier in form of impurities
and defects. In fact, two step tunnelling could, in principle, affect the
conductance for both regimes discussed above. This is due to the unavoidable
presence of defects inside the barrier, even without Si doping
\cite{guerrero2006}, which mix the spin currents and are a source of an
unpolarized current. Therefore, this implies the presence of a finite
characteristic spin-flip time on the defects and on the silicon layer both for
the low and the high Si thickness regimes \cite{Barnas1999}.

\begin{figure}[t]
\includegraphics[width=0.8\linewidth,clip=]{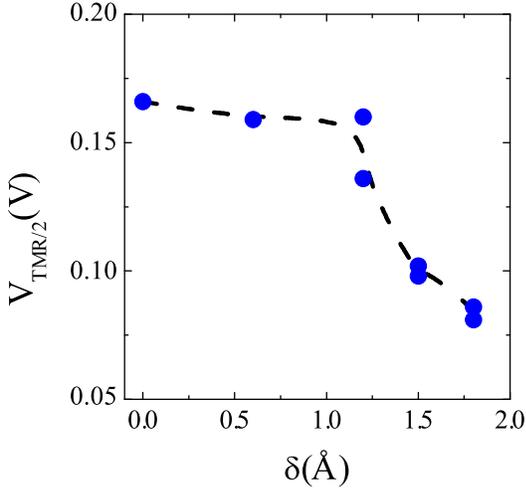}\caption{Voltage needed to
reduce the tunnelling magnetoresistance to $1/2$ of its value at zero bias at
T$\sim$2K. This value quantifies the dependence of the magnetoresistance on
the bias voltage. A low value of V$_{TMR/2}$ means a strong dependence on bias
voltage, while a high value means a weak one. We observe a diminished value of
V$_{TMR/2}$ when the ZBA starts to increase (see Fig. \ref{ZBA} ). }%
\label{TMRhalf}%
\end{figure}

The analysis of bias dependence of TMR which follows further supports our
hypothesis. In order to analyze the bias dependence of TMR as a function of Si
thickness, we have found the voltage needed to suppress a TMR to its half
(zero bias) value (i.e. to TMR(0~V)/2). This parameter, called $\text{V}%
_{\text{TMR}/2}$ is shown in Fig.\ref{TMRhalf}, plotted as a function of the
silicon thickness. Evidently, there is a crossover from a nearly constant
$\text{V}_{\text{TMR}/2}$ regime below $\delta\;$=1.2 \AA , to a strongly
decreasing one $\text{V}_{\text{TMR}/2}$ as a function of $\delta$, for Si
thicknesses above one monolayer. The low Si doping regime with nearly
constant $\text{V}_{\text{TMR}/2}$ proves the presence of two step tunnelling
through localized states, with character and density of states nearly
unchanged up to Si thickness of 1.2~\AA .

When $\delta\geq1.2\AA$ seems that the effective capacitance
corresponding to localized states inside the barrier is reduced, increasing
dramatically the number of states available inside the barrier for tunnelling.
The new transport channels serve as a source of unpolarized current, which
explains the much stronger voltage dependence of TMR for large Si doping.

\section{Theoretical model}

To account for the above discussed experimental features of
electronic transport in tunnel junctions, we calculate now
theoretically transport characteristics as a function of bias
voltage $V$ using the model of rectangular tunneling barrier with
a thin layer of impurities (Si atoms) which create a number of
impurity levels inside the barrier. In the limit of low density of
Si levels, the current is due to direct tunneling in each spin
channel, and the current density can be calculated from the
formula

\begin{align}
j(V)=\frac{e} {4\pi^{2}\hbar} \sum_{\sigma}\int d\varepsilon\int
_{0}^{[2m(\varepsilon-V_{l\sigma})]^{1/2}/\hbar }
k_{l}dk_{l} \left|  t_{\mathbf{k}%
\sigma}\right|  ^{2}\nonumber\\
\times\left(  \frac{\varepsilon-\hbar^{2}k_{l}^{2}/2m-V_{r\sigma}}
{\varepsilon-\hbar^{2}k_{l}^{2}/2m-V_{l\sigma}}\right)  ^{1/2} \left[
f(\varepsilon)-f(\varepsilon-eV)\right], \label{t1}%
\end{align}
where $\mathbf{k}_{l}$ is the in-plane wavevector component of an
electron incident on the barrier, $t_{\mathbf{k}\sigma}$ is the
transmission amplitude for an electron with wavevector
$\mathbf{k}$ and spin $\sigma$, while $V_{l\sigma}$ and
$V_{r\sigma}$ are the spin-dependent energy band edges on the left
and right sides of the junction, respectively, which depend on the
applied voltage as $V_{l\sigma} =V_{l\sigma0}+eV/2$ and
$V_{r}=V_{r\sigma0}-eV/2$ (here $V_{l\sigma0}$ and $V_{r\sigma0}$
are the corresponding band edges at zero voltage). Apart from
this, the integration in Eq.~(\ref{t1}) is over the electron
energy $\epsilon$, and $f(\varepsilon)$ is the Fermi-Dirac distribution
function.

To calculate the tunneling probability for a nonzero voltage
applied to the system we use a semiclassical approximation for the
wavefunction inside the barrier with a slope of potential. This is
justified in case when the variation of the barrier height $V_{0}$ is
small at the electron wavelength $\lambda\sim\hbar
/(2mV_{0})^{1/2}$, which restricts the bias voltage to  $|eV|\ll m^{1/2}V_{0}%
^{3/2}L/\hbar$, where $L$ is the barrier width.


\subsection{Impurity-mediated tunneling}

As the density of impurities grows, an additional mechanism of
tunneling through the impurity levels inside the barrier becomes
more effective than the direct tunneling. In the frame of the
Larkin-Matveev \cite{Larkin87} model, the resonant tunneling
through the structure is described by the transition probabilities
$w_{\mathbf{kp}}$. More specifically, $w_{\mathbf{kp}}$ is the
probability of transition from the state described by the
wavevector $\mathbf{p}$ on the left side of the barrier to the
state corresponding to the wavevector $\mathbf{k}$ on the right
side (from now on we drop the spin index $\sigma$ referring to the
tunneling in different spin channels), and is given by the formula
\begin{equation}
w_{\mathbf{kp}}=\frac{2\pi}{\hbar}\left\vert \sum_{\mathrm{i}}\frac{T_{\mathbf{k}%
\mathrm{i}}T_{\mathrm{i}\mathbf{p}}}{\varepsilon_{\mathbf{p}}-\varepsilon_{\mathrm{i}}+i\Gamma_{\mathrm{i}}%
}\right\vert ^{2}\delta(\varepsilon_{\mathbf{p}}-\varepsilon_{\mathbf{k}%
}),\label{t2}%
\end{equation}
where $T_{\mathbf{p}\mathrm{i}}$ and $T_{\mathrm{i}\mathbf{k}}$ are the matrix
elements for transitions between the states of the corresponding
leads and of the $\mathrm{i}$-th impurity, whereas $\Gamma_{\mathrm{i}}$ is the
width of the impurity level associated with tunneling from the
localized level through the barrier. The sum in Eq.~(\ref{t2})
runs over all impurities. The $i$ factor is the imaginary unit.

An important point is that the matrix elements $T_{\mathbf{p}\mathrm{i}}$
and $T_{\mathrm{i}\mathbf{k}}$ include a phase factor depending on the
location of the impurity inside the barrier,
\begin{equation}
T_{\mathbf{k}\mathrm{i}}=\frac{1}{S^{1/2}}\;e^{-i\mathbf{k}_{l}\cdot\mathbf{R}%
_{\mathrm{i}}}\;e^{k_z(z_{\mathrm{i}}-L/2)}\;\frac{\hbar^{2}}{m}\,(2\pi\kappa)^{1/2}, \label{t3}%
\end{equation}%
\begin{equation}
T_{\mathrm{i}\mathbf{p}}=\frac{1}{S^{1/2}}\;e^{i\mathbf{p}_{l}\cdot\mathbf{R}_{\mathrm{i}}%
}\;e^{-p_{z}(z_{\mathrm{i}}+L/2)}\;\frac{\hbar^{2}}{m}\,(2\pi\kappa)^{1/2}, \label{t4}%
\end{equation}
where $S$ is the junction area, $\mathbf{R}_{\mathrm{i}}$ and $z_{\mathrm{i}}$ are
the in-plane and out-of-plane components of the $\mathrm{i}$-th impurity
position, $\mathbf{k}\equiv (\mathbf{k}_l,ik_z)$ (and similarly
for $\mathbf{p}$), while $\kappa$ is the inverse localization
length of the impurity wavefunction. We assume that the impurities
are randomly distributed in the plane, i.e. $\mathbf{R}_{\mathrm{i}}$ is a
random variable, whereas $z_{\mathrm{i}}$ is the same for all impurities.



Assuming that the energy level $\varepsilon_{\mathrm{i}}$ and the level
width $\Gamma_{\mathrm{i}}$
do not depend on the position $\mathbf{R}_{\mathrm{i}}$, and averaging over $\mathbf{R}%
_{\mathrm{i}}$ in the plane, we obtain the following formula for the
electric current:
\begin{align}
j(V)=\frac{2\pi
e}{S\hbar}\sum_{\sigma}\sum_{\mathbf{k,p}}\frac{\overline
{\left\vert \sum_{\mathrm{i}}T_{\mathbf{k}\mathrm{i}}T_{\mathrm{i}\mathbf{p}}\right\vert
^{2}}
}{(\varepsilon_{\mathbf{k}}-\varepsilon_{\mathrm{i}})^{2}+\Gamma_{\mathrm{i}}^{2}}\nonumber\\
\times\delta(\varepsilon_{\mathbf{k}}-\varepsilon_{\mathbf{p}})\left[
f(\varepsilon_{\mathbf{k} })-f(\varepsilon_{\mathbf{k}}-eV)\right]. \label{t5}%
\end{align}
After calculating the average $\overline {\left\vert
\sum_{\mathrm{i}}T_{\mathbf{k}\mathrm{i}}T_{\mathrm{i}\mathbf{p}}\right\vert ^{2}}$, one
finds that the current density $j$ in the
resonance-impurity channel consists of two terms, and can be written as%
\begin{align}
j(V)=\frac{2\pi
e}{\hbar}\;\sum_{\sigma}\sum_{\mathbf{k,p}}\frac{n\left\vert
T_{\mathbf{k}\mathrm{i}}\right\vert ^{2}\left\vert T_{\mathrm{i}\mathbf{p}}\right\vert ^{2}%
}{(\varepsilon_{\mathbf{k}}-\varepsilon_{\mathrm{i}})^{2}+\Gamma_{\mathrm{i}}^{2}}\left[
1+n\,\delta
(\mathbf{k}_{l}-\mathbf{p}_{l})\right] \nonumber\\
\times\delta(\varepsilon_{\mathbf{k}}-\varepsilon_{\mathbf{p}})\left[
f(\varepsilon_{\mathbf{k}})-f(\varepsilon_{\mathbf{k}}-eV)\right]
.\hskip0.3cm\label{t6}%
\end{align}
The first term in Eq.~(\ref{t6}) is linear in the 2D impurity
density $n$ and describes the transitions through completely
isolated single levels. Such transitions do not conserve the
in-plane components of  $\mathbf{p}$ and $\mathbf{k}$. The second
term in Eq.(\ref{t6}) is nonlinear in $n$ and describes the
electron transitions through the impurity plane. For such
transitions the corresponding in-plane components of electron
wavevectors are conserved.

\begin{figure}[ptb]
\vspace*{-0.5cm}
\includegraphics[width=0.7\linewidth]{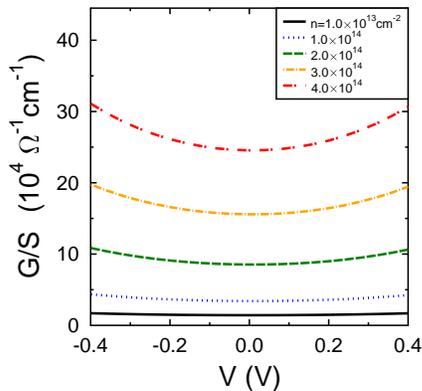}
\vspace*{-0.5cm}\caption{Total conductance per unit area, $G/S$,
calculated as a function of the bias voltage for indicated areal
impurity concentrations. The other parameters are described in the
text.
}%
\label{totalG}%
\end{figure}

\begin{figure}[ptb]
\vspace*{-0.5cm}
\includegraphics[width=0.7\linewidth]{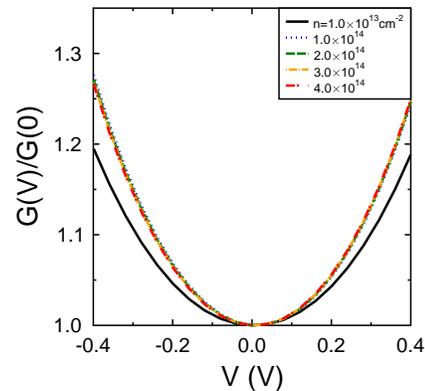}
\vspace*{-0.5cm}\caption{Reduced conductance, $G(V)/G(0)$, as a
function of bias voltage $V$ for indicated density of Si atoms. The other parameters as in Fig.~\ref{totalG}}%
\label{redG}%
\end{figure}

In our calculations we include all three channels. The total
conductance per unit area, $G/S$,
is presented in Fig.~\ref{totalG} for parallel magnetic
configuration and for indicated impurity concentrations. The width
of the tunnel barrier is taken as $L=1.2$~nm, and the Si atoms are
located within the plane of $z_{\mathrm{i}}=-0.4$~nm, measured from the
center of the barrier.
%
The energy structure corresponds to the majority and minority
bands in Co, $E_{F\downarrow}=4.5$~eV and $E_{F\uparrow}=0.66$~eV,
respectively. The height of the barrier is assumed to be
$V_{0}=1$~eV. In turn, in Fig.~\ref{redG} we show the reduced
conductance, $G(V)/G(V=0)$.

It should be noted that the conductance as well as the reduced
conductance are slightly \textit{asymmetric} with respect to bias
reversal, see Fig.~\ref{totalG} and Fig.~\ref{redG}. This
asymmetry is here associated with tunneling through single
impurity levels which are located asymmetrically within the
barrier. In our calculations we assumed that the impurities are
located in the barrier  close to the interface between the barrier
and one of the electrodes -- like in experiments discussed above.
Such an asymmetry of impurity position with respect to the center
of tunnel barrier leads to the asymmetry of the conductance
$G(V)$.

\subsection{Role of Coulomb interaction}

When the density of impurity levels $n$ grows, the conductance
calculated within the model described above and including both
direct and impurity-mediated tunneling increases monotonically
(see Fig.~\ref{totalG}). The experimental data
(Fig.~\ref{res300K&2K}), however,  reveal a rather sharp increase
of the resistance to much higher values when the density $n$
crosses a critical value $n_{cr}$ corresponding to nearly complete
filling of one atomic layer with Si atoms. The results for TMR
presented in Figs.~\ref{TMRTemp} and \ref{TMRhalf} also show that
the physics of tunneling is substantially different for the
density of Si atoms corresponding to complete filling of the
plane.

We assume that the physical reason of such transition is related
to a dramatic increase of the role of Coulomb interaction in the
conductance through the Si levels. This effect can be also
described by the decrease of the number of active levels, which
are able to transmit electrons through the barrier. Indeed, if
some Si atoms form a small cluster, then the cluster of several
atoms acts as a single level for transmission because the Coulomb
interaction prevents two or more electrons to occupy the same
cluster. Thus we can assume that Eq.~(\ref{t6}) describes the
conductance as a function of the density of effective levels,
$n\rightarrow n_{ef}$, corresponding to the number of impurity
clusters. As the density of Si atoms approaches the critical value
$n_{cr}$, the value of $n_{ef}$ decreases rapidly. If $n>n_{cr}$
and the density of Si atoms keeps growing then it corresponds to
increasing thickness of the layer completely filled with the Si
atoms. In such a case the Coulomb interaction is suppressed as
there are no small clusters anymore, and the conductance can be
described using a model of three-layer structure with well-defined
properties of each of the layers. One can expect that the
properties of the Si layer in such a structure are close to those
of layered amorphous Si.

It should be emphasized that the direct tunneling is also suppressed in the
vicinity of $n\sim n_{cr}$. This is related to the Coulomb repulsion of
electrons transmitted through the barrier from the charged impurity clusters,
so that the electrons can tunnel through the barrier only in those areas,
which are free from the impurity clusters or islands. One can describe this by
a local increase of the tunneling barrier in the areas filled with clusters.
This effectively leads to an increase of the average tunneling barrier.

\begin{figure}[ptb]
\vspace*{-0.5cm}
\includegraphics[width=0.7\linewidth]{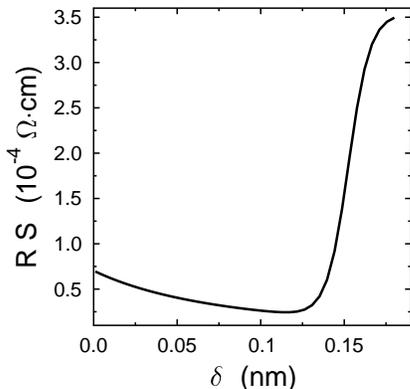}
\vspace*{-0.5cm}\caption{Variation of the resistance per unit square as a
function of Si layer thickness. }%
\label{ResvsDelta}%
\end{figure}



Using the above described ideas we have calculated the resistance
as a function of $\delta$, taking into account all three channels
of conductivity as described above, but instead of the number of
impurities $n$ we put in Eqs.~(\ref{t5}),(\ref{t6}) an effective
number of levels $n_{ef}$, which we assumed to change rapidly from
$n_{ef}=n$ at $n<n_{cr}\simeq1.1\times10^{14}$~cm$^{-2}$ to a
constant value $10^{13}$~cm$^{-2}$. We also corrected the
contribution due to direct tunneling making it strongly dependent
on $n$ in the vicinity of the transition point $n=n_{cr}$. More
specifically, we reduced this contribution for $n>n_{cr}$ by
modeling the dependence of the tunnel barrier $V_{0}$ on $n$: for
$n<n_{cr}$ we take $V_{0}=const$ independent of $n$ but for
$n>n_{cr}$ we assume that this value increases by $0.6$~eV, which
corresponds to suppression of the direct tunneling due to the
'screening' from the large impurity clusters within the Si layer.
The results for the resistance as a function of $\delta$ are
presented in Fig.~\ref{ResvsDelta}. As one can note, the
theoretical curve is qualitatively similar to the experimental
one.


Further improvement of  the model can be made by taking into
account the dependence of the impurity density of states $\nu$ on
energy in the vicinity of the Fermi level, $\varepsilon=\mu$.
%
One can assume that the function $\nu(\varepsilon)$ has a minimum near
$\varepsilon=\mu$ in accordance with the shape of the density of states (DOS)
in amorphous Si (see, for example, Refs.~\cite{lee1994,allan1998}). The DOS
profile in the vicinity of the minimum at $\varepsilon=\mu$ in amorphous a-Si
can be approximately presented as a sum of the DOS tails related to the
conduction and valence bands, $\nu(\varepsilon)\simeq\nu_{0}\left(
e^{-(\varepsilon-\mu+\Delta)/\varepsilon_{0}}+e^{(\varepsilon-\mu
-\Delta)/\varepsilon_{0}}\right)  $, where $\Delta\simeq0.25$~eV$^{-1}%
$atom$^{-1}$, and $\varepsilon_{0}\simeq100$~meV. In the vicinity of the
minimum at $\varepsilon=0$ one can use the parabolic approximation
\begin{equation}
\nu(\varepsilon)\simeq\nu(\mu)\left(  1+\frac{(\varepsilon-\mu)^{2}%
}{2\varepsilon_{0}^{2}}\right)  \label{t7}%
\end{equation}
where $\nu(\mu)=2\nu_{0}e^{-\Delta/\varepsilon_{0}}$.

\begin{figure}[ptb]
\vspace*{-0.5cm} \includegraphics[width=0.7\linewidth]{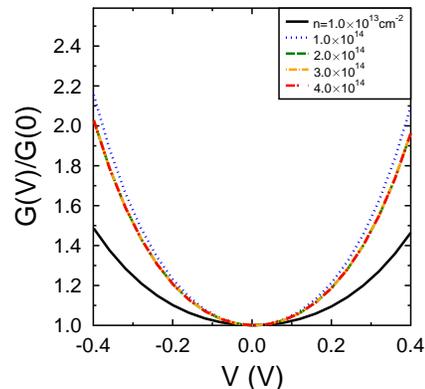}
\vspace*{-0.5cm}\caption{Reduced conductance in the model with Coulomb
repulsion and the DOS of amorphous Si layer taken into account. }%
\label{Gredamorph}%
\end{figure}

After substituting the constant DOS $\nu(\mu)$ by the approximate
function (\ref{t7}) with $\varepsilon_{0}=100$~meV, we finally
find the dependence presented in Fig.~\ref{Gredamorph}. The
calculated dependence shown in Fig.~\ref{Gredamorph} is in
reasonable qualitative agreement with the experimental curves
Fig.~\ref{GV},a.

\section{Discussion}

Several attempts have been made previously trying to understand the possible
role of Coulomb blockade in the tunnelling current and magnetoresistance for
ferromagnetic leads contacting a quantum dot \cite{Barnas1998,Barnas2000}.
These works predict an oscillation in TMR(V) with a period given by the
charging voltage. Experimentally, for tunnelling through an array of dots, the
oscillatory behaviour has been reported only for the magnetic tunnelling
junctions with a barrier doped with cobalt nanoparticles. Those junctions had
rather small ferromagnetic electrodes area A=0.5$\times$0.5 $\mu$m$^{2}$
with the conduction almost completely blocked at low bias \cite{CB2002}, and
TMR oscillating with the period predicted theoretically. Other experiments,
which also studied spin dependent electron transport through arrays of dots
doping the barrier of magnetic tunnel junctions, have reached conclusions in
respect to conductance similar to those reported here, i.e observation of a
non-oscillating increase of the conductivity with bias, when Coulomb blockade
is suppressed by the applied voltage
\cite{CB1996,CBFert1997,Coulombblock1998,CB2003}.

The above cited reports employed different devices in order to study spin
dependent tunnelling through a medium controlled by Coulomb blockade, and two
of them specifically Ref. \cite{CBFert1997} and Ref. \cite{CB2002} used qualitatively
similar magnetic tunnel junction devices. In these papers a granular film,
consisting of nanometer size cobalt particles (with radius close to $\sim$2.5
nm) was embedded in a matrix of aluminium oxide. This array of Co dots was
deposited on top of the aluminium oxide barrier (2.7nm in the first case
\cite{CBFert1997} and 1-2 nm in the second one \cite{CB2002})and was covered
by the second aluminium oxide barrier. The top barrier was different in the
studies mentioned above. While the first paper \cite{CBFert1997} used a $\sim
$1.5 nm thick secondary barrier, the later work did not use any alumina
barrier to cover the Co nanoparticles, which probably produced an uncontrolled
secondary barrier. Another difference is related to the junctions area. In the
first study a rather big (4.5$\times$10$^{-2}$ mm$^{2}$) area MTJ was used,
while the second work studied Coulomb blockade controlled spin dependent
tunnelling in MTJ's with much smaller area junction. All different studies of
the spin dependent transport in MTJ's with nanoparticle doped barrier, with
the exception of Yakushiji et al., \cite{CB2002}, reported a staircase
dependence of the IVs presumably due to single electron charging effects, and
did not show any conductance oscillation.

Although our samples have a silicon $\delta$ layer inside the barrier, instead
of a magnetic $\delta$ layer introduced in the previous reports, the observed
behaviour of the conductance vs. voltage is in general similar to the data
reported for spin tunnelling through arrays of magnetic nanoparticles
\cite{CBFert1997}, with exception of the dependence of TMR on bias voltage
which is smoother (\textrm{V}$_{\text{TMR}/2}\sim$0.5 V, \cite{CBFert1997})
than in our samples.


\begin{figure}[ptb]
\includegraphics[width=0.8\linewidth,clip=]{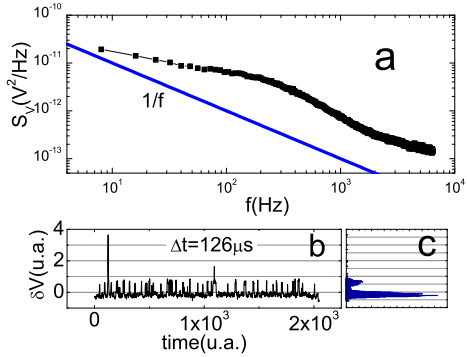}\caption{Typical random
telegraph noise process. The plot (a) shows the power spectrum of the process.
It is dominated by a Lorentzian added to a 1/f noise background (straight line
in the graph). The graph (b) of the figure shows a typical time series. The
two states fluctuation is clearly seen in the histogram (shown in (c)). The
histogram shows the number of counts at a certain bin of voltage. Two peaks
correspond to two states of different conductance. }
\label{RTN}
\end{figure}

A further confirmation of the role of the Coulomb blockade in our samples could be observed in the noise at low frequency in the studied samples. Whereas for low Si doping ($\delta<$
1.2~\AA ), the power spectrum at low temperatures is \textquotedblright
white\textquotedblright\ (i.e. nearly frequency independent) and corresponds
well to the shot noise expected for direct tunnelling in a tunnel junction (or
for two-step tunnelling through strongly asymmetric barriers), for $\delta\geq$1.5~\AA a random telegraph noise (RTN), Fig. \ref{RTN}, contribution becomes evident for bias voltages above a critical value. The appearance of RTN for $\delta\geq$1.2~\AA might be understood as a consequence of the suppression of Coulomb blockade.

RTN has been previously reported for nonmagnetic tunnel junctions and some
other devices such as field effect transistors or quantum dots connected to
metallic leads \cite{Peters1999}. As to the tunnel junctions, RTN has been
usually attributed to resistance fluctuations due to a single or few
fluctuators \cite{Kogan}. The RTN was usually found for rather small area
(below 1$\mu$m$^{2}$) junctions and low temperatures, because in this case the
tunnel resistance is controlled by a few fluctuating defects, providing
two-state fluctuations of the resistance. In the case of rather large tunnel junctions, as in the present study, the
observation of RTN could not be described by the above models, involving
direct influence of single or few defects fluctuations on the resistance. A
non-uniform current distribution, induced by pin-holes, which are a source of
\textquotedblright hot spots\textquotedblright\ just before the MTJs are
broken down by the intensity of current, would neither explain the observed
voltage dependence of the RTN. Indeed, our experimental data, particularly the
current-voltage characteristics and the temperature dependence of the
conductance show absence of pinholes and the above mentioned
\textquotedblright hot spots\textquotedblright.

As discussed above, for thick enough silicon layers $\delta\geq$
1.2~\AA \thinspace\ the effective capacitance of the Si dots becomes small
enough to break down the Coulomb blockade above a certain bias voltage. This
increases the electron population of the island due to the two steps
tunnelling events and enhances the tunnelling conductance.

\begin{figure}
\includegraphics[width=\linewidth]{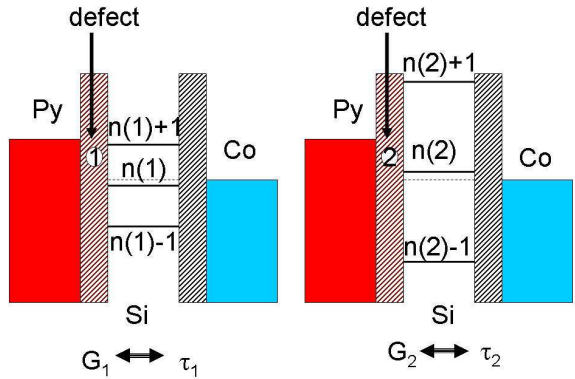}\caption{Schematic explanation of the noise observed at low temperatures. The presence of traps in our barrier modifies the levels inside our dot. This effect leads to two resistance states depending on what is the state of the defect. This effect in visible at low temperatures because only few processes are activated, hence the noise is RTN type as is clearly seen in Fig. \ref {RTN}.}%
\label{model}%
\end{figure}

It is evident that in the system under study the capacitance of the Si dots
and, correspondingly, the Si dots population, should be distributed over the
MTJ area, providing a possible variation of the local tunnelling current as a
function of the spatial coordinate. In addition, the two-level systems
situated close to the Si dots seem to introduce also time dependent
fluctuations or RTN in the tunnelling current through these dots. The
unavoidable dependence of the tunnelling current on the coordinate may enhance
the effective contribution to the overall conductance from only a few
fluctuators, resulting in a noise contribution additional to 1/$f$ due to
effective \textquotedblright amplification\textquotedblright\ of some
conductance fluctuations responsible for RTN from these few defects. This is represented schematically in the Fig. \ref{model}: the change in the charge in the defect yields two states with different tunneling rates ($\tau$), hence different conductance. The origin of the different times lies in the different levels inside the dots, which are determined by the capacitance, thus by the local environment. Of course, in our tunnel junction there are many defects, but at low temperatures there are only several of them active because the trapping-detrapping process uses to be thermally activated.

\section{Conclusions}

To conclude, we have carried out an extensive study of electron transport in
Co$\vert$Al$_{2}$O$_{3}$$\vert$Py magnetic tunnel junctions asymmetrically doped with Si. Our experimental
data suggest that the observed behaviours of conductance vs Si
doping are closely related, clearly indicating a suppression of the Coulomb
blockade regime for Si layer thicknesses above about one monolayer. Although
no staircase behavior of the conductance was observed for the regime of
suppressed Coulomb blockade, as in some other systems \cite{CB2002}, the
observed behavior of tunnelling conductance is rather similar to the one
reported long time ago by Giaever \cite{Giaever69}, which was successfully
explained as due to the presence of a large amount of particles inside the
barrier. For a fixed bias voltage applied in equilibrium, these particles
might have very different electron population, even though their capacitances
are characterized by rather narrow size distributions, giving rise to some
distribution of the maximum threshold voltage suppressing Coulomb blockade
$\pm$V$_{ch}$. This leads to the observed zero bias anomaly in our samples. On
the other hand, the variation of the electron population in each particle may
explain the suppression of the predicted \cite{Barnas1998} Coulomb
oscillations, which could be present only for a constant equilibrium electron
population in Si particles across the junction area. One of the ways to reach
a more uniform electron population in the Si particles could be the reduction
of the area of the junctions as in Ref. \cite{CB2002}, implying tunnelling through
a smaller array of nanoparticles with a size distribution narrower than in the
present case.

\begin{acknowledgments}
This work was supported by funds from the Spanish Comunidad
de Madrid (Grant No. P2009/MAT-1726) and Spanish
MICINN  (Grants No. MAT2006-07196, No. MAT2009-
10139, Consolider Grant No. CSD2007-00010). As a part of
the European Science Foundation EUROCORES Programme
Grant No. 05-FONE-FP-010-SPINTRA, work was supported
by funds from the Spanish MEC (MAT2006-28183-E), Polish
Ministry of Science and Higher Education as a research
project in years 2006–2009, and the EC Sixth Framework
Programme, under Contract No. ERAS-CT-2003-980409.
The work was also supported by ESF-AQDJJ programme,
FCT Grant PTDC/FIS/70843/2006 in Portugal, and by the
Polish Ministry of Science and Higher Education as a research
project in years 2007–2010 (V.K.D.). Work in MIT is
supported by NSF Grant No. DMR-0504158 and ONR Grant
No. N00014-06-1-0235.
\end{acknowledgments}



\end{document}